\pdfoutput=1
\documentclass[aps,prd,reprint,superscriptaddress,longbibliography]{revtex4-1}

\usepackage{graphicx}
\usepackage{amsmath}
\usepackage{amssymb}
\usepackage{amsfonts}
\usepackage{dcolumn}
\usepackage{dsfont}
\usepackage{latexsym}
\usepackage{rotating}
\usepackage{color}
\usepackage{latexsym}
\usepackage{bbm}
\usepackage{subfigure}
\usepackage{float}
\usepackage{epsfig}
\usepackage{psfrag}
\usepackage{natbib, hyperref}
\usepackage{bm}
\usepackage{amsthm}
\usepackage{eucal}
\usepackage{mathrsfs}
\usepackage{url}
\usepackage{braket}
\usepackage{ulem}

\usepackage{calligra}
\usepackage[T1]{fontenc}
\usepackage[utf8]{inputenc}
\usepackage{newunicodechar}
\usepackage{marvosym}
\usepackage[absolute]{textpos}
\usepackage[cal=cm]{mathalfa}

\usepackage[resetlabels]{multibib}  
\newcites{Supp}{Supplementary~References}   

\usepackage{color} 

\usepackage{hyperref}
\hypersetup{
colorlinks=true,final=true,
        linkcolor=blue,
        citecolor=blue,
        filecolor=blue,
        urlcolor=blue,
}

\begin{document}

\title{Electrical control of Majorana Bound States Using Magnetic Stripes}


\author{
Narayan Mohanta$^{1}$,~Tong Zhou$^{2}$,~Jun-Wen Xu$^{3}$,~Jong E. Han$^{2}$,\\
Andrew D. Kent$^{3}$,~Javad Shabani$^{3}$,~Igor \v{Z}uti\'c$^{2}$,~Alex Matos-Abiague
}

\affiliation{
\textit{Department of Physics \& Astronomy, Wayne State University, Detroit, MI 48201, USA}\\
$^{2}$\textit{Department of Physics, University at Buffalo, State University of New York, Buffalo, NY 14260, USA}\\
$^{3}$\textit{Center for Quantum Phenomena, Department of Physics, New York University, NY 10003, USA}
}

\begin{abstract}
A hybrid semiconductor-superconductor nanowire on the top of a magnetic film in the stripe phase experiences a magnetic texture from the underlying fringing fields. The Zeeman interaction with the highly inhomogeneous magnetic textures generates a large synthetic spin-orbit coupling. We show that this platform can support the formation of Majorana bound states (MBS) localized at the ends of the nanowire. The transition to the topological superconducting phase not only depends on the nanowire parameters and stripe size but also on the relative orientation of the stripes with respect to the nanowire axis. Topological phase transitions with the corresponding emergence or destruction of MBS can be induced by reorienting the stripes or shifting their position, which can be achieved by passing a charge current through the magnetic film or by applying electrically-controlled strain to it. The proposed platform removes the need for external magnetic fields and offers a non-invasive electrical tuning of MBS with the perturbation (current or strain) acting only on thee magnetic film.
\end{abstract}
           
\maketitle

\section{Introduction}
Impressive experimental advances have made possible the realization and detection of Majorana bound states (MBS) in systems that under certain conditions can be driven into the topological superconducting phase~\cite{Mourik2012:S,Das2012:NP,Rokhinson2012:NP,Finck2013:PRL,Zhang2018:N,Nadj-Perge2014:S,Pawlak2016:NPJQI,Deng2016:S,Feldman2017:NPhys}. Because of their non-Abelian statistics and topological protection, MBS are promising for the implementation of fault-tolerant topological quantum computing~\cite{Kitaev2003:AP,Nayak2008:RMP}. Such an implementation requires not only the creation but also the manipulations of MBS in order to realize braiding and fusion operations~\cite{Kitaev2001:PU,Kitaev2003:AP,Nayak2008:RMP,Sau2010:PRL,Aasen2016:PRX,Schuffelgen2019:SSE}. The majority of proposals relies on systems with strong Rashba SOC and the use of an external magnetic field as a tuning tool~\cite{Fu2008:PRL,Lutchyn2010:PRL,Oreg2010:PRL,Grosfeld2011:PNAS,Pientka2013:NJP,Cook2011:PRB,Narayan2014:EPL,Brouwer2011:PRB,Braunecker2013:PRL}. However, MBS can still be realized without the need for Rashba SOC nor external magnetic fields if appropriate magnetic textures are used. It is known that magnetic textures can generate both Zeeman and synthetic spin-orbit coupling~\cite{Korenman1977:PRB,Tatara1997:PRL,Bruno2004:PRL,Jia2010:PRB}, which together with superconductivity provide the basic ingredients for the formation of MBS. Therefore, there has been an increasing interest in the use of magnetic textures for the generation of MBS~\cite{Jelena:PRL2012,Nadj-Perge2013:PRB,Li2016:NComm,Klinovaja2013:PRL,Kjaergaard2012:PRB,Braunecker2010:PRB,Vazifeh2013:PRL,Marra2017:PRB,Yang2016:PRB,Gungordu2018:PRB,Xu2015:PRL,Sun2016:PRL,Kim2015:PRB}. In particular, the use of locally tunable magnetic fringing fields can enable the realization of both braiding and fusion operations~\cite{Fatin2016:PRL,Matos-Abiague2017:SSC,Tong_PRB2019}. The suitability of using synthetic SOC generated by magnetic fringing fields for the creation of MBS in a superconducting carbon nanotube has recently been experimentally demonstrated~\cite{Desjardins2019:NM,Yazdani2019:NM}. 

In this article, we show that in spite of its non-helical character, the fringing field generated by experimentally created magnetic stripes in a Co/Pt multilayer thin film can support the formation of MBS in a semiconductor nanowire (NW) with proximity-induced superconductivity. The controllability of the stripes orientation and geometry by means of a charge current~\cite{Lahtinen_SCiRep2012,Jiang2017:PhysRep,Wang2018:APL}, the magnetoelectric effect~\cite{Chung2008:APL}, or by applying strain~\cite{Nong2018:IEEETM} allows for electrically tuning topological phase transitions in the NW. Our work reveals opportunities to extend the concept of proximitized materials~\cite{Zutic_MaterToday2019} and their applications.
\begin{figure}[htb!]
\begin{minipage}[c]{0.47\textwidth}
\centering
\vspace{-0.2cm}
\epsfig{file=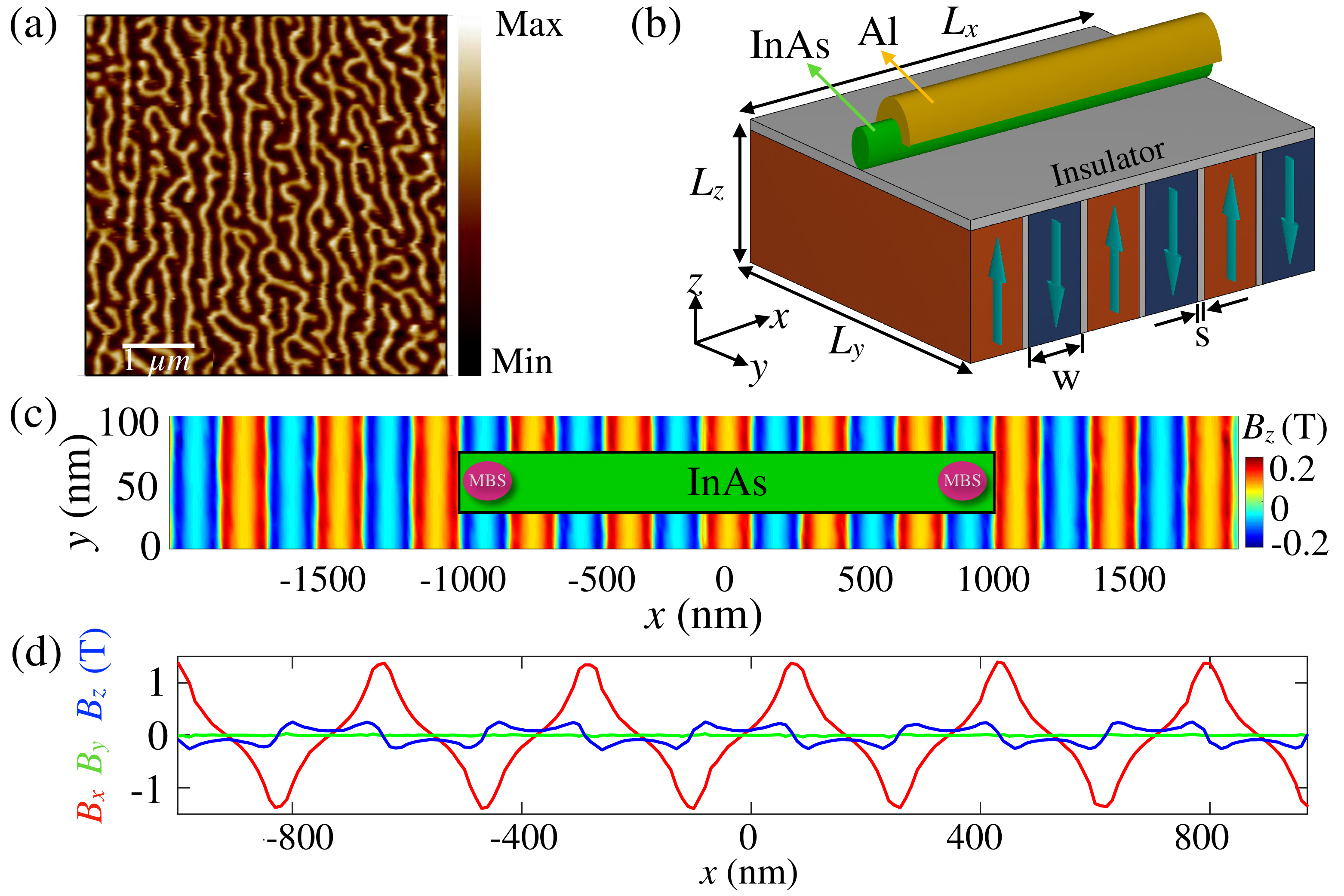,trim=0.0in 0.0in 0.0in 0.0in,clip=false, width=85mm}
\caption{(a) Magnetic force microscopy image of the magnetic stripe phase formed in a Co/Pt multilayer thin film. (b) Schematic picture of the proposed set-up, consisting of an InAs-Al nanowire placed on the top of magnetic stripes. The nanowire is separated from the magnetic film by a thin insulating spacer (gray layer). (c) Micromagnetic simulation of the $z$-component of the fringing field in the absence of magnetic disorder, as experienced by the carriers in the nanowire for a $5$ nm thick insulating spacer and stripe domains of width (W) $160$ nm, thickness ($L_z$) $40$ nm, and domain-wall width ($S$) 20 nm. The color represents the out-of-plane component ($B_z$). The saturation magnetization ($M_s$) is $1.7\times10^{6}$~A/m. The nanowire is positioned at the middle, rectangular region (green) with dimensions $2~\mu$m$\times50$~nm. (d) Variation of the three components of the fringing field with distance ($x$) along the length of the nanowire.}
\label{fig1}
\end{minipage}
\end{figure}

\section{Characterization of magnetic stripes}
The magnetic films consist of Co/Pt multilayers deposited on oxidized silicon wafers in an ultra-high vacuum physical vapor deposition system (Kurt Lesker PVD-12) by magnetron sputtering with the substrate at ambient temperature. The film consists of 20 Pt|Co repeats on a thin Ta seed layer capped with Pt, specifically, Ta(3nm)|[Pt(0.3nm)|Co(0.6nm)]x20|Pt(0.3nm). The domain images are obtained with a magnetic force microscope (MFM) at room temperature, shown in Fig.~\ref{fig1}(a). An in-plane magnetic field of about 1~T is initially applied to align the stripes along the field direction, as shown in Ref.~\onlinecite{Kent2001:JPCM}. The formed stripes are stable and survive for a long time after the external magnetic field is switched off. In this study, as detailed below, we explore the possibility of using the fringing fields generated by the stable stripes in the absence of an external magnetic field for inducing topological phase transitions in a semiconductor NW with proximity-induced superconductivity.
 
 \section{Theoretical model}
A schematic of the proposed set-up is shown in Fig.~\ref{fig1}(b). Superconductivity is induced in the InAs NW by proximity to the Al half-covering. Although any semiconducting NW with high $g$-factor could be suitable, we consider InAs in which epitaxial growth and proximity-induced superconductivity have been demonstrated~\cite{Javad2017:PRB,Javad2018:PRB,Mayer2018:arXiv}. The NW is separated from the magnetic film by an insulator spacer (gray layer). We use realistic micromagnetic modeling of the magnetic textures using the finite-element method in COMSOL~\cite{COMSOL}. The magnetic fringing fields are simulated using stripe domains similar to the ones formed in experimentally realizable Co/Pt multilayer thin films. Due to the complexity of the problem, we consider a simplified, two-dimensional (2D) version of the actual NW. The $z$-component of the fringing field ($B_z$) in the plane of the NW is shown, as a function of position, in Fig.~\ref{fig1}(c), where the green region represents the InAs NW with dimensions 2~$\mu$m$\times$50 nm. In our stripe geometry the fringing field is sizable at the bottom of the wire and falls off rapidly with vertical distance. Therefore, the formation of the MBS is expected to be restricted towards the bottom of the nanowire and an effective 2D model represents a reasonable first-approach to capture the essential physics in this geometry. For the numerical simulations we assumed a 5 nm thick insulating spacer and a film section of size ($L_x=$4 $\mu$m)$\times$($L_y=$100 nm)$\times$($L_z$=40 nm) and saturation magnetization $M_s$=$1.7\times10^{6}$~A/m~\cite{Tsymbal:2012}. The stripes have a width W=160~nm and are separated by domain-walls of thickness S=20~nm. Changes in the fringing field components ($B_x$, $B_y$, and $B_z$) along the $x$ axis are displayed in Fig.~\ref{fig1}(d). The maximum amplitude of the fringing field lies below the critical field of about 1.9~T experimentally measured in InAs/Al NWs~\cite{Krogstrup2015:NM}. We note that although $B_y$ is negligibly small and both $B_x$ and $B_z$ exhibit an oscillatory behavior, the $x$-component of the fringing field largely dominates. The resulting fringing field is therefore quite different from the helical-like and skyrmion-like textures considered in previous investigations of MBS~\cite{Yang2016:PRB,Gungordu2018:PRB}. Therefore, it is somehow surprising that in spite of having a non-helical character, the textures illustrated in Fig.~\ref{fig1}(d) can also support the formation of MBS, as shown below. While any magnetic gradient creates a synthetic SOC, not every synthetic SOC necessarily leads to the formation of MBS. In the current magnetic-stripes geometry, the $x$-component of the fringing field is much larger than the $z$-component, distinguishing it from other non-helical spin textures~\cite{Fatin2016:PRL,Tong_PRB2019} that give rise to the formation of MBS. This leads to the general question of which kind of magnetic textures, and therefore, which kind of synthetic SOC are compatible with the generation of MBS. An approximate topological condition as a function of slowly varying magnetic textures has been previously derived~\cite{Fatin2016:PRL,Matos-Abiague2017:SSC} but a general answer to the question remains open. 

The InAs NW is proximity-coupled to superconducting Al and described by the following Hamiltonian, 
\begin{align}
\mathcal{H}=&\sum_{i,\sigma} (4t-\mu) c_{i\sigma}^{\dagger}c_{i\sigma}-t\sum_{\langle ij \rangle,\sigma}(c_{i\sigma}^{\dagger}c_{j\sigma}+H.c.) \nonumber \\
&+\sum_{i} (\Delta\; c_{i\uparrow}^{\dagger} c_{i\downarrow}^{\dagger}+H.c.) -\frac{g^*\mu_{_B}}{2} \sum_{i} (\mathbf{B}_i\cdot \boldsymbol{\sigma})_{\sigma \sigma'} c_{i\sigma}^{\dagger}c_{i\sigma'} \nonumber \\
&-\frac{i\alpha}{2a}\sum_{\langle ij \rangle,\sigma,\sigma'} (\boldsymbol{\sigma} \times \boldsymbol{d}_{ij})_{\sigma\sigma'}^{z} c_{i\sigma}^{\dagger}c_{j\sigma'},
\label{Ham}
\end{align} 
\noindent where $c_{i\sigma}^{\dagger}$ ($c_{i\sigma}$) is the fermionic creation (annihilation) operator at site $i$ with spin $\sigma$, $t$=$\hbar^2/(2m^*a^2)$ is the hopping energy, $m^*$ is the effective mass of electrons, $a$ is the unit lattice spacing of the underlying square lattice, $\Delta$ is the proximity-induced superconducting gap, and $\mu$ is the chemical potential measured from the lowest energy state of the semiconductor wire. The real-space information of the fringing field $\mathbf{B}_i$ is contained in the Zeeman interaction (fourth term), where  $g^*$ is the effective $g$-factor of electrons, $\mu_{_B}$ is the Bohr magneton and $\boldsymbol{\sigma}$ represents the Pauli spin matrices. The last term represents the Rashba SOC originating from the broken structure inversion symmetry~\cite{Zutic2004:RMP}, with $\alpha$ being its strength and $\boldsymbol{d}_{ij}$ the unit vector between sites $i$ and $j$. We use $m^*$=$0.026\;m_0$ (effective mass for InAs), $a$=$10$~nm, $\Delta$=$0.2$~meV, and unless otherwise specified, $\alpha$=$10$~meVnm, $g^*$=$15$~\cite{Antipov2018:PRX}. 

The Hamiltonian~(\ref{Ham}) is solved by exact diagonalization after performing a transformation of the fermionic operators to the Bogoliubov-de Gennes (BdG) basis: $c_{i\sigma}$=$\sum_{n} u_{i\sigma}^n\gamma_{n}+v_{i\sigma}^{n*}\gamma_{n}^{\dagger}$, where $u_{i\sigma }^n$ ($v_{i\sigma}^n$) is the BdG quasiparticle (quasihole) amplitude, and $\gamma_{n}^{\dagger}$ ($\gamma_{n}$) is a fermionic creation (annihilation) operator of a BdG quasiparticle or quasihole in the $n^{\text{th}}$ energy eigenstate. 

\begin{figure}[t]
\centering
\begin{minipage}[c]{0.47\textwidth}
\centering
\epsfig{file=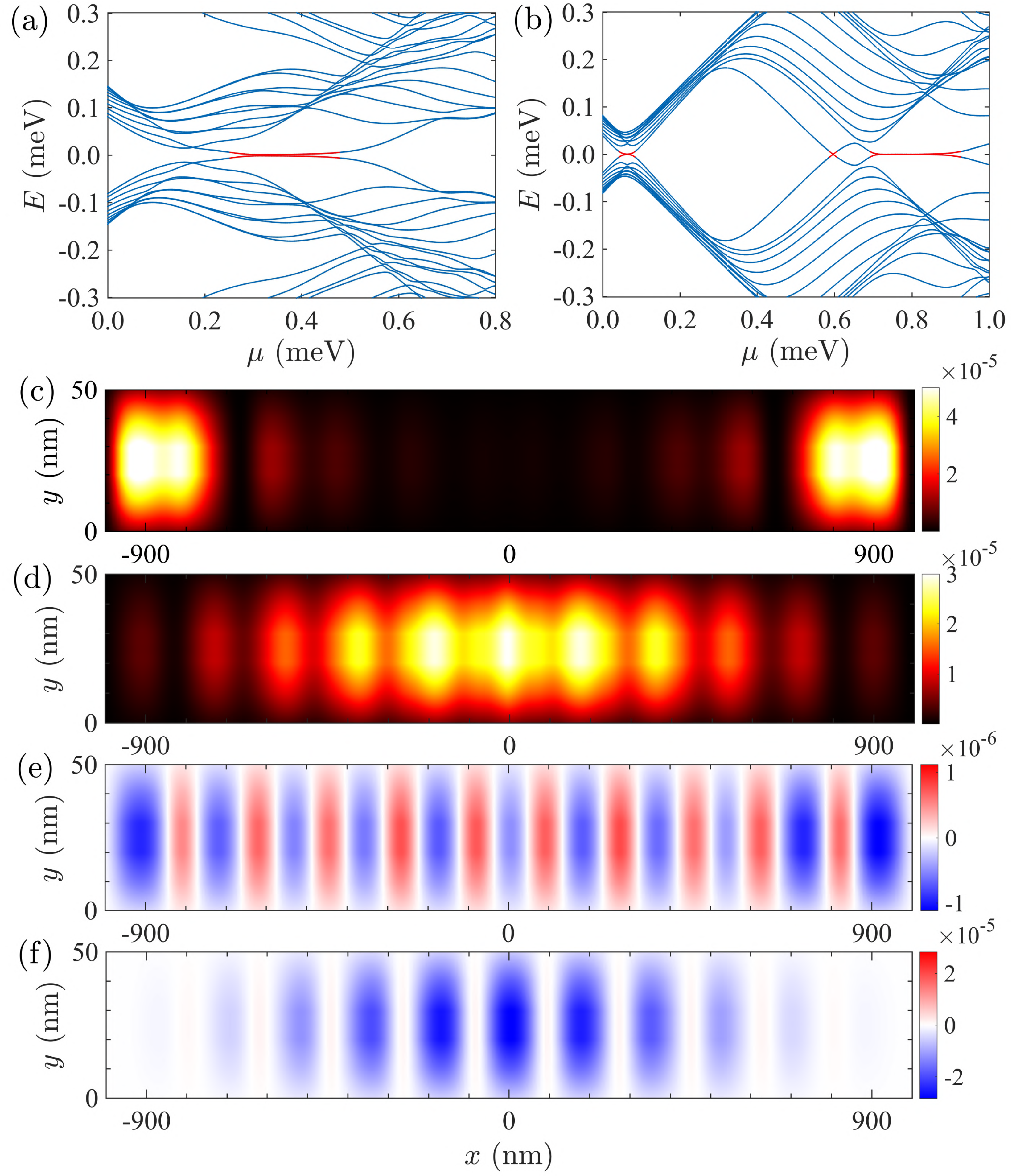,trim=0.0in 0.0in 0.0in 0.0in,clip=false, width=85mm}
\caption{Energy spectrum, plotted as a function of the chemical potential ($\mu$), for stripes with period $\lambda=360$~nm for (a) $g^*$=$15$ and (b) $g^*$=$30$. The red dots show the zero-energy MBS. Real-space profile of the probability density $|\Psi_i|^2$, in units of 1/nm$^2$, corresponding to the lowest-energy state at (c) $\mu$=$0.35$~meV  and (d) $\mu$=$0.65$~meV for $g^*$=$15$. Real-space profile of the charge density $\rho_i$, in units of $e$/nm$^2$, corresponding to the lowest-energy state at (e) $\mu$=$0.35$~meV  and (f) $\mu$=$0.65$~meV for $g^*$=$15$.}
\label{fig2}
\end{minipage}
\end{figure}

\section{Results}
\subsection{Formation of Majorana bound states}
When the chemical potential $\mu$ is increased, the energy gap closes within a range of $\mu$ and zero-energy MBS emerge, as shown in Fig.~\ref{fig2}(a), where the computed low-energy spectrum is displayed. Within this range (plotted in red), other energy levels are shifted away from zero energy, creating a \textit{minigap} that protects the MBS. Near $\mu$=$0.08$~meV, the lowest pair of energy levels come closer to zero energy but do not meet each other to produce robust MBS. To illustrate the effect of increasing the $g$-factor, the $\mu$ dependence of the low-energy spectrum for $g^*$=$30$ [i.e., twice the InAs $g$-factor used in Fig.~\ref{fig2}(a)] is shown in Fig.~\ref{fig2}(b). In practice, the effective $g$-factor can be enhanced in semiconducting quantum wells using alloys such as InAs$_{x}$Sb$_{1-x}$~\cite{Svensson2012:PRB}. Magnetically doped semiconductors such as (In,Mn)As can reach $g$-factor values above 120~\cite{Zutic2004:RMP} or even 300 in (Cd,Mn)Te~\cite{Betthausen2012:S}. A comparison between Figs.~\ref{fig2}(a) and (b) indicates that for larger $g$-factors, the MBS are not only more robust but they appear within multiple ranges of $\mu$. The existence of multiple topological regions originates from the mixing of different subbands when the $g$-factor is large enough. 

To visualize the real-space localization of the MBS, we plot the probability density $|\Psi^2_i|$=$\sum_{\sigma} (|u_{i\sigma}^1|^2+|v_{i\sigma}^1|^2)$ corresponding to the lowest energy eigenstates at $\mu$=0.35~meV (with existing MBS) and $\mu$=0.65~meV (without any MBS) for $g^*$=$15$, as shown in Fig.~\ref{fig2}(c) and \ref{fig2}(d), respectively. The states are sharply localized near the ends of the NW for $\mu$=0.35~meV, and inside the NW for $\mu$=0.65~meV, clearly distinguishing between the MBS and the non-MBS states. The probability densities of the two MBS at the two NW ends decay exponentially with distance and create a finite overlap, resulting in a finite splitting of the lowest energy pair when the length of the NW is small (not shown here). We also plot the local charge density $\rho_i$=$e\sum_{\sigma} (|u_{i\sigma}^1|^2-|v_{i\sigma}^1|^2)$ for these two values of the chemical potentials in Fig.~\ref{fig2}(e) and \ref{fig2}(f). The charge density exhibits an oscillating behavior along the  NW length. The reduction by an order of magnitude in the amplitude of the charge density for $\mu$=0.35~meV, compared to that for $\mu$=0.65~meV, is a reflection of the nearly zero-charge character of the MBS compared to the sizable charge of finite-energy states. Local charge measurements, avoiding direct contact to the NW, could be used to detect MBS~\cite{BenShach2015:PRB,Scharf2015:PRB}, as an alternative to the quantized zero-bias signature in the tunneling conductance~\cite{Sengupta2001:PRB,Scharf2015:PRB,BenShach2015:PRB,Narayan2018:PRB,PhysRevB.96.014513,DasSarma2012:PRB,Rokhinson2012:NP,Pikulin2012:NJP,Kells2012:PRB,Bagrets2012:PRL}. 

\begin{figure}[t]
\centering
\begin{minipage}[c]{0.47\textwidth}
\centering
\epsfig{file=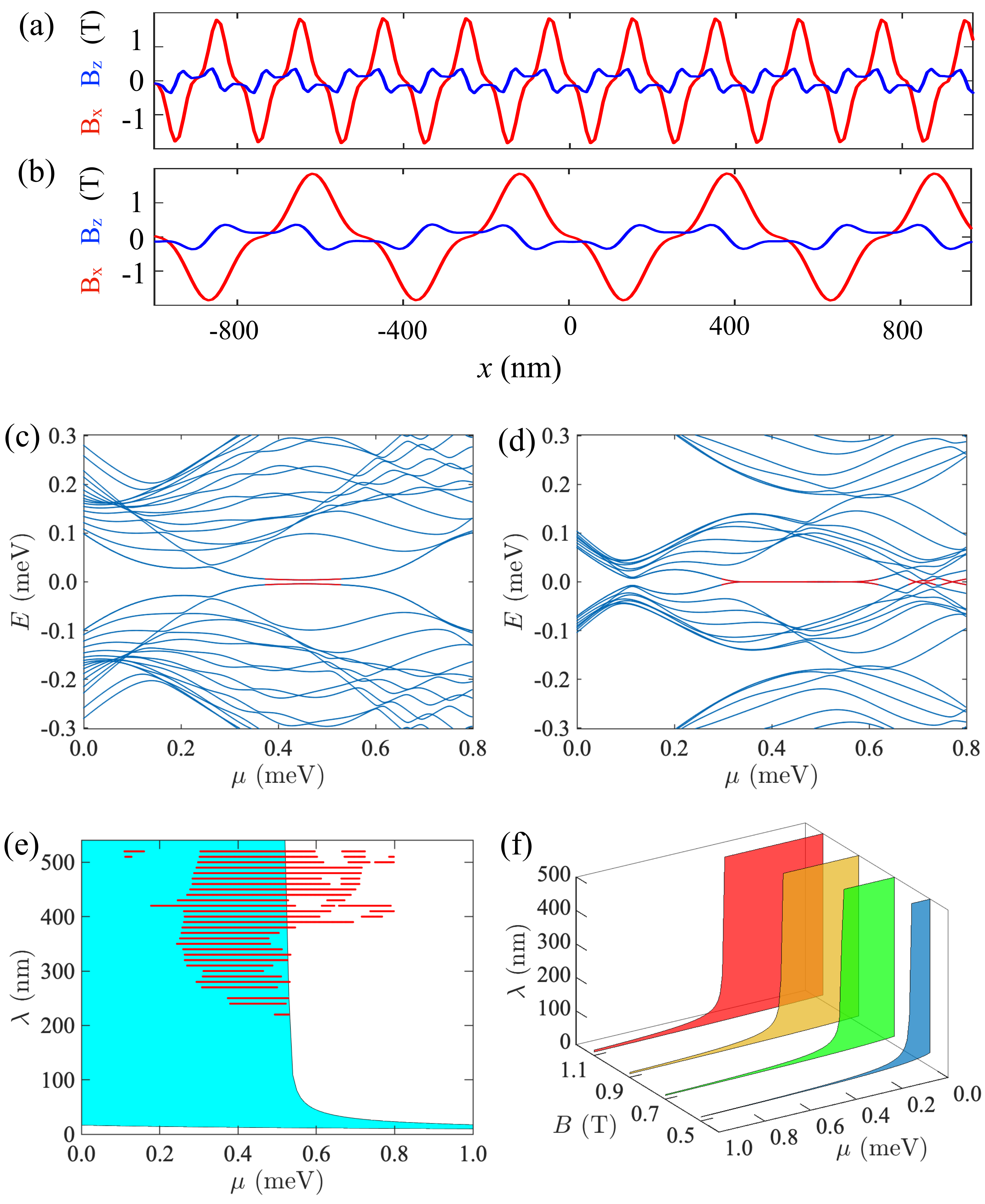,trim=0.0in 0.0in 0.0in 0.0in,clip=false, width=85mm}
\caption{Two components ($B_x$ and $B_z$) of the fringing field, plotted as a function of the distance ($x$) along the length of the nanowire for two different stripe periods: (a) $\lambda$=$250$~nm and (b) $\lambda$=$500$~nm. Low-energy spectrum as a function of the chemical potential ($\mu$) for (c) $\lambda$=$250$~nm and (d) $\lambda$=$500$~nm. (e) Phase diagram in the $\mu-\lambda$ plane, showing the topological superconducting regime (in cyan), calculated for a one-dimensional (1D) nanowire in the presence of a helical texture with maximum field amplitude $B$=$1.3$~T. The red lines obtained from numerically solving the BdG equation represent some of the $\mu$ values for which MBS exist in the nanowire under the fringing field of magnetic stripes with different values of the period $\lambda$. (f) Phase diagrams for a 1D nanowire in a helical texture with different values of field amplitudes: $B$=$0.5$~T (blue), $B$=$0.7$~T (green), $B$=$0.9$~T (yellow), $B$=$1.1$~T (red). }
\label{fig3}
\end{minipage}
\end{figure}

In the case of a helical texture, the strength of the generated synthetic SOC is given by the helix wave number $q=2\pi/\lambda$ with $\lambda$ denoting the period of the helix. In our platform the fringing field, although of non-helical character, is characterized  by the stripe period $\lambda=2{\rm W}+{\rm S}$, as illustrated in Fig.~\ref{fig1}(b). It is therefore relevant to investigate the topological phase transition of the proximitized NW as a function of $\lambda$. The position dependence of the $x$ and $z$-components of the fringing fields generated by a stripe phase with periods $\lambda$=250~nm and $\lambda$=500~nm are shown in Figs.~\ref{fig3}(a) and(b), while the corresponding low-energy spectrum as a function of $\mu$ is displayed in Figs.~\ref{fig3}(c) and (d), respectively. We note that both $\lambda$=$250$~nm and $\lambda$=$500$~nm leads to the formation of MBS for appropriate values of $\mu$. However, for longer periods, the MBS emerge over a larger range of $\mu$. Furthermore, for  $\lambda$=$500$~nm, multiple topological regions develop as the result of subbands mixing.
\begin{figure}[t]
\centering
\begin{minipage}[c]{0.47\textwidth}
\centering
\epsfig{file=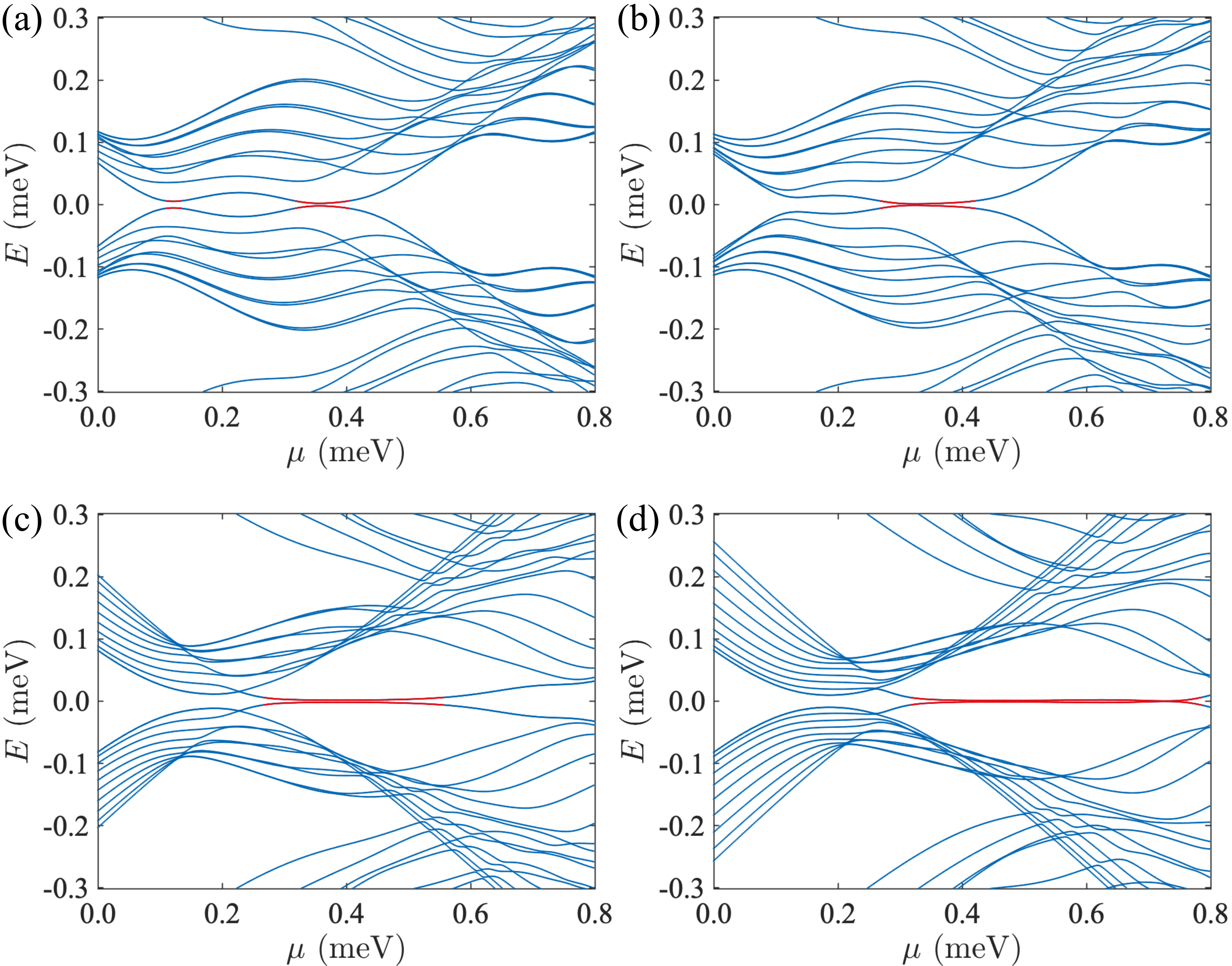,trim=0.0in 0.0in 0.0in 0.0in,clip=false, width=85mm}
\caption{Low-energy spectrum as a function of the chemical potential for different strengths of Rashba spin-orbit coupling (a) $\alpha$=0, (b) $\alpha$=5~meVnm, (c) $\alpha$=15~meV\AA, and (d) $\alpha$=20~meV\AA. All other parameters are the same as in Fig.~2. }
\label{soc}
\end{minipage}
\end{figure}

In order to better understand the effects of both $\lambda$ and $\mu$ on the realization of the topological superconducting phase, we first consider, for the sake of comparison, the case of a one-dimensional (1D) NW under a helical magnetic texture. By defining $g^*\mu_B B/2$=$J$, where $B$ is the amplitude of the helical magnetic field, the topological condition can be written as $J^2>(\mu-\eta)^2+\Delta^2$, where $\eta=\hbar^2q^2/8m^*$ and $q=2\pi/\lambda$. This imposes boundaries on the values of $\lambda$ for which the topological phase can be reached,
\begin{align}
\sqrt{\frac{\pi^2t}{\mu+\sqrt{J^2-\Delta^2}}} < \frac{\lambda}{a} < \sqrt{\frac{\pi^2t}{\mu-\sqrt{J^2-\Delta^2}}},
\label{condition2}
\end{align}
where the upper bound applies only when $\mu>\sqrt{J^2-\Delta^2}$. The corresponding phase diagram in the $\mu$-$\lambda$ plane is shown in Fig.~\ref{fig3}(e), where the cyan region represents the topological superconducting regime specified by Eq.(\ref{condition2}) at $B$=$1.3$~T. To compare this phase diagram with the solutions of Hamiltonian~(\ref{Ham}) for our platform, we compute the energy spectrum for several values of $\lambda$, ranging from $150$~nm to $520$~nm, and plot the values of $\mu$ at which MBS appear [red segments in Fig.~\ref{fig3}(e)] in the same plot. Interestingly, in spite of the non-helical character of the magnetic texture generated by the stripes, many of the topological regions obtained for the set-up in Fig.~\ref{fig1}(b) lie within the topological region of the helical texture on a 1D NW. Therefore, the phase diagram for the magnetic helix case still can be used for a rough estimation of the system parameters leading to the formation of MBS in the 2D-NW/stripes set-up, specially for $\mu \lesssim 0.5$~meV. For larger values of $\mu$, the occupancy of multiple subbands in the 2D NW under the non-helical fringing field of the stripes leads to re-entrance into the topological regime and the emergence of MBS beyond the topological region of the 1D-NW/magnetic helix structure (note the multiple red dots outside the cyan region in Fig.~\ref{fig3}(e)]. A phase diagram including the magnetic helix amplitude, in addition to $\lambda$ and $\mu$ is shown in Fig.~\ref{fig3}(f) for the 1D-NW/magnetic helix structure. As explained above, we can use this phase diagram to draft some conclusions about the system parameters required to drive the proposed platform into the topological regime. Thus, Fig.~\ref{fig3}(f) suggests that it is desirable to have larger fringing fields, wider stripes, and a smaller chemical potential to favor the formation of robust MBS.

In Fig.~\ref{soc}(a)-(d), we show the low-energy spectra for different values of Rashba SOC strength $\alpha$. Although increasing the Rashba SOC enhances the range of $\mu$ leading to the formation of MBS, its presence is not a requirement for a transition to the topological superconducting phase. Indeed, as shown in Fig.~\ref{soc}(a), the synthetic SOC resulting from the fringing field is enough to induce the formation of MBS, even in the absence of Rashba SOC.
\begin{figure}[t]
\centering
\begin{minipage}[c]{0.47\textwidth}
\centering
\epsfig{file=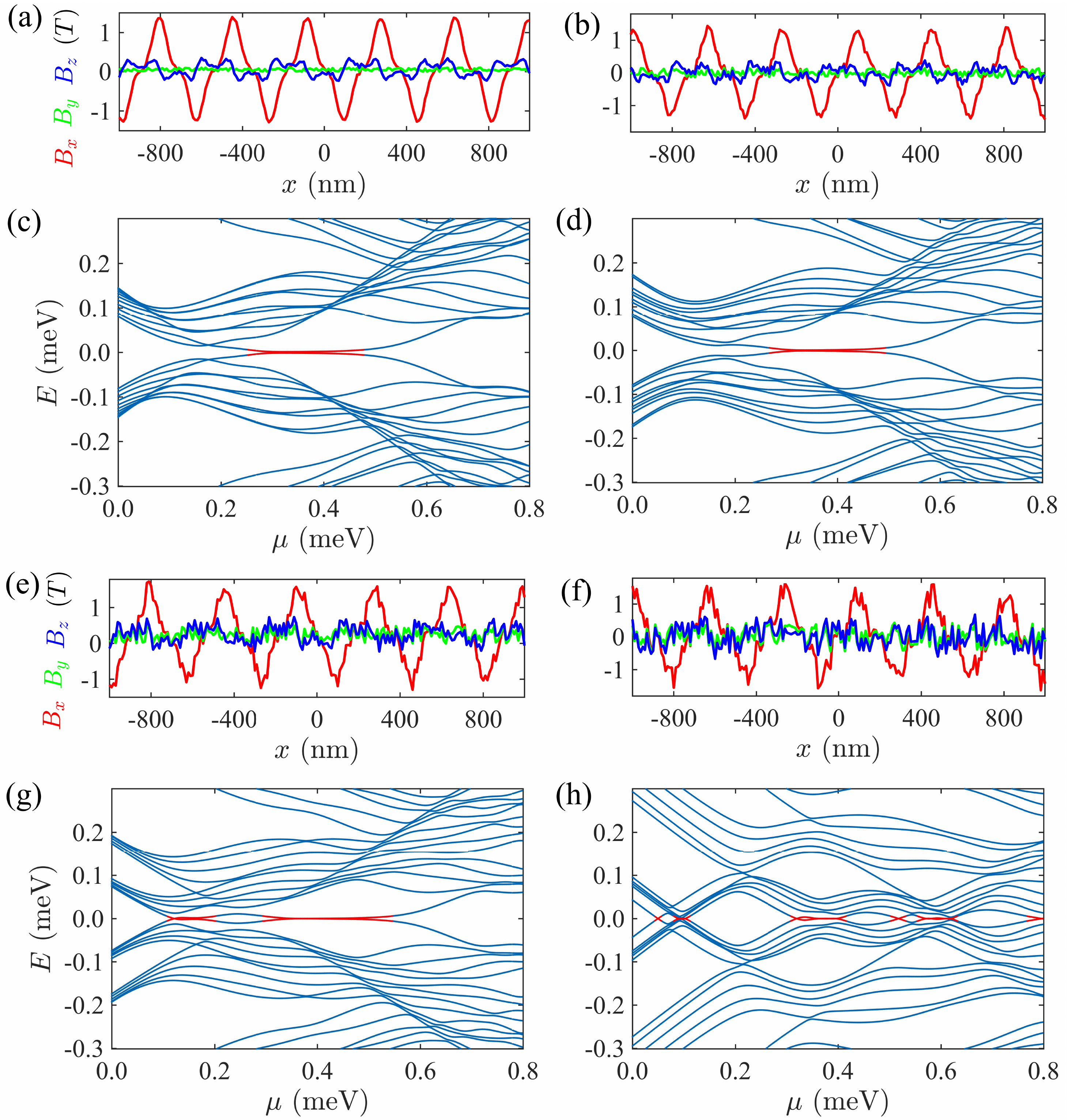,trim=0.0in 0.0in 0.0in 0.0in,clip=false, width=85mm}
\caption{Three components of the fringing field for the homogeneous case and three different values of the strength ($V_d$) of the stripe inhomogeneity are plotted in (a) $V_d$=0, (b) $V_d$=0.3~T, (e) $V_d$=0.6~T and (f) $V_d$=0.9~T, respectively. Low-energy spectrum as a function of the chemical potential for (c) $V_d$=0, (d) $V_d$=0.3~T, (g) $V_d$=0.6~T, and (h) $V_d$=0.9~T. Parameters: $g^*$=$15$, $\lambda$=$360$~nm. All other parameters are the same as in Fig.~2. }
\label{fig5}
\end{minipage}
\end{figure}

\subsection{Effects of magnetic stripes inhomogeneity}
We have also investigated the stability of MBS in the presence of random fluctuations of the fringing fields generated by the magnetic stripes. Due to imperfections and roughness of the magnetic stripes, the fringing field can exhibit fluctuations on a length scale smaller than the stripe width. In order to explore the effects of such magnetic disorder on the MBS, we considered random variations of the fringing field within the range $[-V_d/2, V_d/2]$ around the calculated values of the three components of the field (\textit{viz.}, $B_x$, $B_y$ and $B_z$). The results, shown in Fig.~\ref{fig5}(a)-(d) for different values of $V_d$, indicate that MBS are quite robust against magnetic disorder, as long as the fluctuation amplitude remains smaller than the maximum field amplitude. The robustness of MBS against magnetic disorder has also been investigated for the case of a superconducting carbon nanotube with synthetic SOC~\cite{Desjardins2019:NM}. A previous study reveals that in the highly disordered situation, Majorana edge modes are critically extended and beyond a disorder threshold value, determined by the localization length of the Majorana states, the Majorana states collapse into Anderson localized states in the bulk, resulting in a topological Anderson insulating state~\cite{Habibi2018:PRB,Samuel2018:PRB}. For the present stripe case, such a disorder threshold will depend on the period ($\lambda$) of the stripe.

\begin{figure}[t]
\centering
\begin{minipage}[c]{0.47\textwidth}
\centering
\epsfig{file=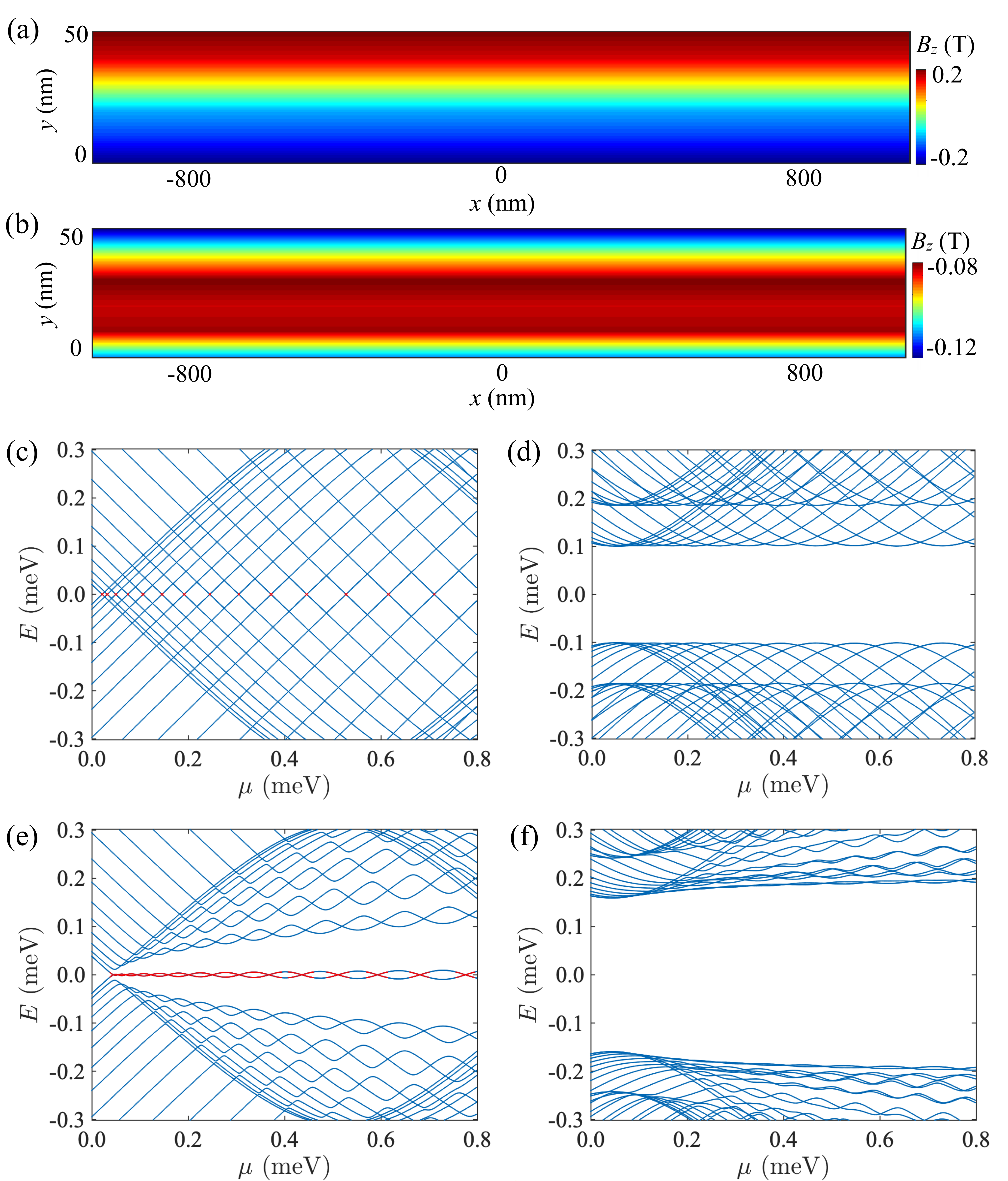,trim=0.0in 0.0in 0.0in 0.0in,clip=false, width=85mm}
\caption{(a) Spatial profile of the out-of-plane component, $B_z$, of the fringing field generated by stripes parallel to a nanowire located on the top of the domain wall between two adjacent stripes. (b) same as in (a) but for a nanowire located on the top of a single stripe. (c) and (d) Low-energy spectrum as a function of the chemical potential, $\mu$, for the cases (a) and (b), respectively in the absence of Rashba SOC (\textit{i.e.} $\alpha$=$0$). (e) and (f) Low-energy spectrum for the cases (a) and (b), with a finite Rashba SOC of strength $\alpha$=$10$~meVnm. Parameters: $\lambda$=$360$~nm, $g^*$=$15$. }
\label{fig6}
\end{minipage}
\end{figure}

\subsection{Tunability of the Majorana bound states}
A particularly attractive functionality of the proposed platform for switching of topological phase transitions relies on the possibility of changing the magnetic stripes orientation by passing a current through the Co/Pt multilayer thin film~\cite{Kooy1960:Philips,Kent2001:JPCM,Lahtinen_SCiRep2012,Jiang2017:PhysRep,Wang2018:APL} and reorienting the stripes from being perpendicularly to being parallel to the NW. Alternatively, the stripe orientation and geometry can also be changed by applying strain to the magnetic film-\cite{Nong2018:IEEETM}. The strain can be electrically controlled by using a piezoelectric substrate.The stripes reorientation modifies the form of the fringing fields experienced by the carriers in the NW and, eventually, affect the topological superconducting phase. Two relevant scenarios might occur when the stripes are parallel to the NW: (i) the NW is located on the top of the domain wall between two adjacent stripes and (ii) the NW lies on the top of a single stripe. The position dependence of the $z$-component of the fringing fields in the NW region for cases (i) and (ii) is shown in Figs.~\ref{fig6}(a) and (b), respectively. In both cases, the fringing field does not change along the NW. 

Let us first discuss the case with $\alpha=0$ (i.e., absence of Rashba SOC). The low-energy spectra corresponding to the cases (i) [Fig.~\ref{fig6}(a)] and (ii) [Fig.~\ref{fig6}(b)], are shown in Figs.~\ref{fig6}(c) and (d), respectively. In both cases the synthetic SOC only couples spin to the $y$-component of the carrier momenta and no MBS are expected to occur. The absence of MBS for when the stripes lie parallel to the NW is clearly manifested in the low-energy spectra shown in Figs.~\ref{fig6}(c) and (d). If the wire lies on the top of a domain wall [case(i)], the Zeeman energy due to the fringing field exceeds the superconducting gap $\Delta$ in most of the NW region, causing the gap to close at certain values of the chemical potential [see Fig.~\ref{fig6}(c)]. In case (ii), however, the fringing field is considerably weaker and the gap prevails over the complete range of chemical potentials [see Fig.~\ref{fig6}(d)]. The absence of MBS when the stripes are parallel to the NW [Fig.~\ref{fig6}(c) and (d)] contrasts with the formation of MBS in the topological superconducting phase when the stripes are in the perpendicular direction [Fig.~\ref{fig6}(a)]. Therefore, the formation of MBS can be manipulated by changing the orientation of the magnetic stripes.

The discussion above remains valid for finite but small $\alpha$. However, for larger Rashba SOC MBS emerge when the wire lies on top of a domain wall, as shown in Fig.~\ref{fig6}(e). Interestingly, MBS are still absent when the wire is located on the top of a stripe [see Fig.~\ref{fig6}(f)]. Therefore, in the case of sizable Rashba SOC, one does not need to reorient the stripes. Instead, positioning the wire parallel to the stripes, the MBS switching can simply be achieved by shifting the position of the domain walls. When a domain wall lies underneath the wire MBS form and they are suppressed when the domain wall moves away and the wire lies on top of a stripe. The effective motion of the domain walls can be realized by passing a current through the magnetic film~\cite{Jiang2017:PhysRep,Wang2018:APL,Parkin2008:Science,Uhlig2009:JAP} or, alternatively, by applying strain and changing the width of the stripes~\cite{Nong2018:IEEETM}.

\section{Discussion and summary}	
The results presented above suggest that an electrically-induced shift or reorientation of the magnetic stripes can be used to control the topological phase transition in the NW without the need for an external magnetic field. The electrical perturbation (charge current, electric field, strain) acts on the Co/Pt multilayer film, which is isolated by the spacer from the NW and although electric contacts on the NW may be needed for MBS detection, the electrically-induced topological phase transition occurs in a non-invasive manner, i.e., with no charge transfer.

The detection of the MBS can be performed using tunneling spectroscopy by attaching metallic leads to the NW and by looking for a zero-bias conductance peak (ZBCP)~\cite{Sengupta2001:PRB,Finck2013:PRL,Lee2012:PRL,Deng2012:NL,Liu2012:PRL}. The electrically-induced switching of the stripes orientation offers an additional knob not only for investigating the nature of the ZBCP and its relation to the presence of MBS, but also for exploring their robustness as a function of tunable synthetic SOC. Furthermore, the proposed set-up could provide a proof-of-concept demonstration of the feasibility of using electrically-controlled synthetic SOC for the manipulation of MBS beyond recent experimental advances~\cite{Desjardins2019:NM}, paving the way to more complex platforms with extended tunability for the realization of braiding operations~\cite{Fatin2016:PRL,Matos-Abiague2017:SSC}.

\section*{acknowledgements}
This work is supported by DARPA Grant No.  DP18AP900007, 
US ONR  Grant No. N000141712793 (I. \v{Z}. and A. M.-A.), and the UB Center for Computational Research. This work was performed in part at the Advanced Science Research Center NanoFabrication Facility of the Graduate Center at the City University of New York.


%

\end{document}